# Interaction Design for VR Applications: Understanding Needs for University Curricula

Oloff C. Biermann*
ocbier@cs.ubc.ca
University of British Columbia
Vancouver, British Columbia, Canada

Daniel Ajisafe*
dajisafe@cs.ubc.ca
University of British Columbia
Vancouver, British Columbia, Canada

Dongwook Yoon
yoon@cs.ubc.ca
University of British Columbia
Vancouver, British Columbia, Canada

## ABSTRACT

As virtual reality (VR) is emerging in the tech sector, developers and designers are under pressure to create immersive experiences for their products. However, the current curricula from top institutions focus primarily on technical considerations for building VR applications, missing out on concerns and usability problems specific to VR interaction design. To better understand current needs, we examined the status quo of existing university pedagogies by carrying out a content analysis of undergraduate and graduate courses about VR and related areas offered in the major citadels of learning and conducting interviews with 7 industry experts. Our analysis reveals that the current teaching practices underemphasize design thinking, prototyping, and evaluation skills, while focusing on technical implementation. We recommend VR curricula should emphasize design principles and guidelines, offer training in prototyping and ideation, prioritize practical design exercises while providing industry insights, and encourage students to solve VR design problems beyond the classroom.

## 1 INTRODUCTION

As a growing technology, virtual reality (VR) offers an unlimited set of possible interactions in 3D environments and has become an active area of interest in building immersive consumer products [6, 16]. According to Statista [20], the sales of virtual and augmented reality headsets are expected to reach over 26 million units by 2023, which presents an opportunity for interaction designers to develop more immersive 3D experiences.

However, designing for VR can be quite challenging as new tools evolve rapidly and different modalities emerge [19]. Understanding the fundamental principles of human-centered interaction is key to building real-world VR experiences. Current curricula in institutions are more focused on teaching software and development practices for VR applications than designing for user interactions in VR. This is problematic as designing good immersive experiences requires teaching fundamental design principles and guidelines [8, 18]. There is an increasing need in education for interaction design principles and guidelines for VR. If formal training from learning institutions is fragmented, there is a danger that the next generation of designers will not be equipped to meet the requirements that current and future VR interaction design demands. While the current curriculum covers design education in HCI in general [11, 13, 23, 24]However, it is not clear to what extent these skills are taught within the VR context. Therefore, we ask the following research question: *"To what extent do the current curricula about VR and related areas prepare students to meet the requirements for designing VR applications?"*

In this study, we acknowledge that there are significant interplays between VR and AR design, and some design principles and guidelines are transferable across the two domains. However, there are design problems that are specific to fully immersive environments in VR. Therefore, we streamline our study to specifically focus on fully immersive environments and the unique design challenges they present for future designers to meet demands in the real world. We note that exploring the use of VR for instruction is **not** the goal of this study. Rather, we focus on teaching interaction design education topics for VR in a way that prepares future designers to meet real-world demands.

First, we conducted a wide-scope overview of institutions publishing VR research in some of the top HCI and VR venues, including CHI [5] and IEEE VR [22]. Aside from these venues, we also included top-ranked institutions in the liberal arts. Thus, we carried out an extensive content analysis on the courses from these institutions and extracted key findings related to current pedagogical methods. We defined pedagogical methods as major means of teaching. These were mainly classified as lecture materials, theoretical/reading assignments, and practical exercises. To further explore these findings, we also conducted an interview study with VR industry and academic experts, thereby obtaining a more diverse view of the needs in the current curricula. Finally, we propose a clear set of recommendations to improve the current pedagogy.

Our study ultimately makes three empirical contributions [25]: 1) Identifying the current learning activities in the curricula about VR and the teaching strategies that are used, 2) revealing the needs for VR interaction design in current curricula, and 3) proposing a set of recommendations for an improved curriculum covering VR design.

## 2 RELATED WORKS

Studies on interaction design education for VR and immersive technologies have been relatively limited. In 2004, Burdea conducted a survey of general VR education which found that only 148 universities globally had VR courses [3]. Burdea's survey continued informally until 2008 to eventually include a total of 273 courses [4]. There has not been a more recent survey on this topic since then. In [10], the authors conducted a high-level search of global universities to evaluate assessment methods and learning activities in augmented reality (AR), analyzing the course content in a small subset consisting of 17 courses. Notably, they found that practical projects were the most commonly used learning activity. By contrast, [14] documents the analysis of a single practical project-based VR course. To the best of our knowledge, ours is the first study that performs a survey on education about VR, specifically regarding interaction design topics.

---

*These authors contributed equally to this work.

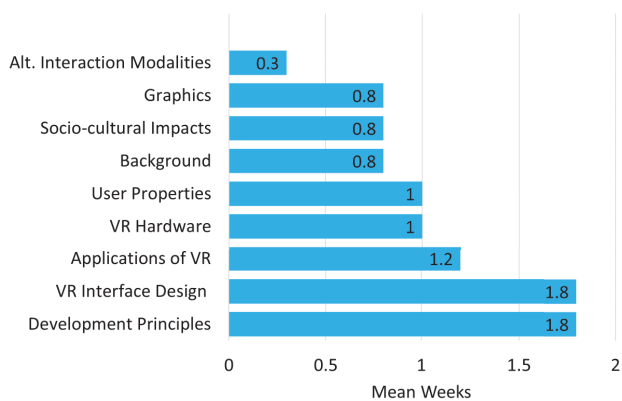

**Figure 1: Mean Weeks Per VR Topic. Topics are: Alternative Interaction Modalities, Graphics, Socio-cultural Impacts of VR, Background of VR, User Properties, VR Hardware, Applications of VR, VR Interface Design, and Development Principles.**

Unlike education for VR, training in interaction design and HCI in general has been somewhat more widely studied [11–13, 23, 24]. In general, some programs and courses emphasize formal iterative design (e.g., human-centered design), while others focus on the creative design process. Thus, the pedagogical methods being used in courses vary significantly and there is a lack of integration between the two design approaches [23]. Considering methodology, the closest work that aligns with our study is [17], which aims to understand the coverage of voice user interface design in HCI pedagogy and advocates for an improved curriculum in traditional learning institutions similar to [2, 9, 15]. Unlike [17], we also incorporate semi-structured interviews to enrich our findings and illustrate how the coverage of interaction design in the current curricula fails to meet VR designers' needs.

## 3 CURRICULA ANALYSIS OF VR COURSES

To establish an understanding of the current state of VR curricula, we conducted an extensive analysis of courses taught at top institutions around the world. We started our analysis by choosing a set of learning institutions (see 3.1.1) and then performing a high-level scoping search (explained in 3.1.2) to identify relevant courses. Finally, we performed an in-depth meta-review of the content of syllabi, as described in 3.1.3.

| VR Course Keywords |
|---|
| Virtual Reality, VR, Augmented Reality, AR, Immersive Environments, HCI, Human-Computer Interaction, Graphics |

**Table 1: List of keywords to identify VR courses**

### 3.1 Methods

*3.1.1 Data Sources for Institution Selection.* We sought to include a wide range of institutions for conducting our initial search, described in 3.1.2. Additionally, we wanted to avoid introducing subjectivity into the selection, as individuals may have different perceptions of what constitutes a "top" institution or department in a particular domain. Therefore, rather than basing our selection on personal opinion, we utilized three data sources for institution selection to equally represent top institutions in liberal arts in general, as well as HCI and VR in particular.

Our first goal was to include top university departments in HCI. Thus, we chose universities with the top average number of CHI publications between 2017 and 2020 [21], inclusive. We used CHI as the community standard for ranking departments in terms of their standing in the field of HCI. Similarly, given our focus on VR, we used the average number of IEEE VR publications between 2017 to 2020 [1], inclusive, for selecting our second set of institutions. Finally, we chose a third set of institutions from the Times Higher Education ranking of top liberal arts institutions [7]. This allowed us to broaden the scope of our search to include institutions that are not prominent in research relating to HCI or VR but are highly ranked in terms of teaching. Our selection approach ultimately allowed us to create a diverse set containing 63 institutions. This is comprised of both public and private institutions. Course data was then extracted from online syllabi and course materials.

*3.1.2 Scoping Search of Courses.* Starting with the final list of institutions, we performed a search to find all undergraduate and graduate courses that covered topics directly related to VR. We continued this search until we reached a total of 50 courses.

An obvious choice was to include courses directly focused on VR in our search. However, we also aimed to include other areas related to VR, such as computer graphics. To this end, we developed several keywords to form a set of uniform selection criteria which can be seen in Table 1. We followed a relatively permissive strategy to avoid excluding courses. Therefore, we included any courses where the course title or course syllabus included at least one of the keywords.

*3.1.3 In-depth Meta-review of Curricula.* After conducting the scoping search to identify our target set of courses, we performed an in-depth meta-review of a randomly sampled subset of 20 courses to analyze the online syllabus and course content of each. We synthesized data on the usage of pedagogical methods and topics within which VR is framed. Our analysis consisted of three stages: 1) a first pass to synthesize common topics and methods, 2) a second pass for extraction and organization, and 3) a third pass for detailed review of course content. This was an iterative, collaborative process with frequent cross-checking between the authors.

| Course Name | University Name |
|---|---|
| C1: Virtual Reality Systems | University of Washington |
| C2: AR/VR Capstone | University of Washington |
| C3: Introduction to AR/VR Application Design | University of Michigan |
| C4: Developing AR/VR Experiences | University of Michigan |
| C5: Virtual Reality | Stanford University |
| C6: 3D User Interfaces | Georgia Institute of Technology |
| C7: Introduction to Mobile XR | University of MaryLand |
| C8: Introduction to VR | University of California (Irvine) |
| C9: Virtual Reality | University of California (Berkeley) |
| C10: Augmented Reality - Medical AR | University of Munich |
| C11: Perception and Learning in Robotics and Augmented Reality | University of Munich |
| C12: VR | University of Illinois at Urbana Champaign |
| C13: VR AR | Purdue University |
| C14: Virtual Reality and Physically-Based Simulation | University of Bremen |
| C15: Introduction to 3D Animation and Virtual Reality | Harvard University |
| C16: Introduction to Human-Computer Interaction | Yale University |
| C17: Emergent Interface Technologies | The University of Chicago |
| C18: Mobile Computing | The University of Chicago |
| C19: 3D User Interfaces and Augmented Reality | Columbia University |
| C20: Interaction Design for Virtual and Augmented Reality | National University of Singapore |

**Table 2: List of analyzed courses and corresponding institutions where they are offered.**

During the first pass, we identified pedagogical methods across all courses by examining the online syllabi. We then synthesized a representative set of pedagogical methods and a taxonomy of main topics (see Appendix A) based on the methods and subtopics we identified within each course. In the second pass, we utilized our established list of pedagogical methods and topic taxonomy to extract course data. We revisited the specific pedagogical methods in each course and then classified them on a per-course basis (summarized in Table 3). We then analyzed the topic coverage for each course based on our taxonomy of main topics and the subtopics covered in these courses. Specifically, we classified the subtopics in each course identified in the first pass for each course according to our taxonomy and recorded the number of weeks devoted to each category. Finally, we calculated the mean number of weeks devoted to each area across all courses. The outcome is presented in Figure 1. On the surface, these results would seem to indicate that practical development and design receive roughly equal coverage within the current curricula. The results also indicate that the subtopic of alternate interaction modalities receives very little coverage. To investigate our outcome further, we conducted a detailed analysis of the content within each course in our third pass. The findings from this analysis are presented in the following section.

| Course | LM | TA | PE |
|---|---|---|---|
| C1 | | | |
| C2 | | | |
| C3 | ✓ | | ✓ |
| C4 | ✓ | | ✓ |
| C5 | | | |
| C6 | ✓ | ✓ | |
| C7 | ✓ | ✓ | |
| C8 | ✓ | ✓ | |
| C9 | ✓ | | |
| C10 | ✓ | | ✓ |
| C11 | | | |
| C12 | ✓ | ✓ | |
| C13 | ✓ | | |
| C14 | ✓ | ✓ | |
| C15 | | | |
| C16 | | | |
| C17 | ✓ | | |
| C18 | | | |
| C19 | | ✓ | |
| C20 | ✓ | | ✓ |

**Table 3: Pedagogical Methods Related to VR Interaction Design in each course where LM = Lecture Material, TA = Theoretical/Reading Assignment, and PE = Practical Exercises. Each column shows whether the corresponding pedagogical method was used.**

## 3.2 Findings

In this section, we expand on the varied findings we extracted from our meta-review and content analysis. Table 2 shows the list of 20 courses reviewed. Based on our content analysis, we identified the following prevalent themes describing our findings about the curricula: 1) Pedagogical methods related to VR interaction design; 2) Coverage for VR interaction design topics; 3) Formulation of current VR interaction design topics.

Three of the main types of pedagogical methods in the courses that are specifically relevant to VR interaction design included 1) lecture materials, 2) theoretical/reading assignments, and 3) practical exercises. The lectures are either instructor-led classes or seminar-based presentations given by guest speakers or a combination of both. The theoretical assignments generally included group discussions, reading responses to relevant papers or additional reading materials related to interaction design and principles. Lastly, the practical exercises were typically weekly or bi-weekly assignments given to the students to implement important concepts discussed in class. Our analysis revealed the following findings:

***Theoretical assignments and practical exercises underemphasize design training.*** Of all 3 pedagogical methods identified, theoretical assignments and practical exercises had the least design-related activity (see Table 3). Out of 20 courses, 12 had no formal theoretical activity on interaction design principles for VR (C1, C2,

C4, C5, C9, C10-C13, and C15-C18). This reveals the lack of theoretical considerations for design principles and guidelines. For example, in C10, no reading material on interaction design was given, and the readings mainly focus on computer vision, and tracking technologies. We found that only 6 courses (C6-C8, C12, C14, and C19) had reading activities related to interaction design in the VR context. For example, C7 provided related readings on designing UX for VR apps. The last 2 courses (C3 and C20) did not disclose their reading activity.

Considering practical exercises, only 4 out of 20 courses (C3, C4, C10, and C20) focused specifically on design thinking and equipping students with the fundamental tools/techniques for building good user experience in VR. For example, C3 provided learning activities on critical design thinking, physical and digital prototyping. Out of 12 assignments for this course, more than half of them are interaction design-centric. However, 8 of the 20 courses (C5-C8, C12-C14, and C17) were largely focused on building VR applications, with the assumption that students would learn good design practices through development, hence overlooking the need for a formal teaching on design guidelines. Finally, 5 of 20 courses (C1, C2, C11, C15, and C18) had no practical activity related to VR interaction design, while C16 and C19 focused on general HCI and AR respectively. Thus, interaction design principles and guidelines are underemphasized in VR curricula.

***Lecture materials do not favor interaction design content.*** Based on the pedagogical methods identified in Table 3, the lecture material remains a core component of teaching. Therefore, we determine the extent of interaction design coverage in the lecture material for each of the courses. We found that 8 of the 20 courses devote no lectures to interaction design principles and guidelines in VR (C1, C2, C5, C11, C15, C16, C18, and C19). For example, C5 was heavily centered on building devices and had no formal lectures on interaction design principles. Although C16 gave lectures on design thinking, research methods, prototyping and tangible user interfaces, their focus was not on VR but HCI in general. We saw that C7 and C17 devote only 1 lecture to VR interaction design, C9, C13, and C14 focus 2 lectures on this area, while C4, C8, and C12 devote 3 lectures to VR design and C10 had 4 lectures covering this topic. Finally, C3 and C20 strongly emphasized the importance of design principles in VR. C20 also invited industry experts to teach students how to build good user experiences for VR, which was lacking in other courses.

***Interaction design topics are either framed as special, evolving, or explorative***. We also aimed at understanding how interaction design topics were being discussed. Both C3 and C4 explained interaction design as an *"evolving"* paradigm. Specifically, C3 extensively discusses new design principles, current development approaches and VR interface evaluation. C6 covers immersive 3D interfaces and interactions but from a graphical user interface perspective. Computer Graphics is required as a pre-requisite to this course, and it makes no distinction between GUI and VR interaction design as part of the course goals. C9 frames its lecture on "Locomotion and Motion Sickness" as a *"special"* topic, while C14 expresses interaction design as an "*explorative* space" in VR.

## 4 EXPERT INTERVIEW STUDY

To understand the perception of what interaction design topics curricula should include and how students can be prepared for future industrial positions, we conducted an expert interview study.

### 4.1 Methods

*4.1.1* ***Participant Selection***. The participants were selected based on background, experience, ease of access, and availability. Recruitment followed a purposive sampling scheme. We interviewed 7 experts (*P*1 to *P*7). Participants are experts either working in industry (*P*1, *P*3, *P*4, *P*6, and *P*7) or academics with industry experience in the VR space (*P*2 and *P*5). They had a minimum of 4 years of experience and a maximum of 19 years of experience, with 7 years on average. Our interviewees were: a senior UX/XR designer, a VR/AR designer, a UX design manager, a senior research scientist, a principal research scientist, an assistant professor, and an associate professor. Unfortunately, all responses to our interview request were from men. We provide additional information on the recruitment process in Appendix B.

*4.1.2* ***Procedure***. Each interview began with a brief introduction of the interview agenda. We generated a list of 8 questions (see Appendix C) to gain each participant's perspective on the extent to which the curricula in VR-related courses prepare students to handle the concerns and usability problems specific to interaction design. Interviews were conducted via an online conferencing platform and transcribed via a commercial transcription service.

First, we inquired from the interviewees as to which topics they considered most important to interaction design. Following this question, we probed further for topics that are not sufficiently covered in curricula, and pedagogical methods that could be applied to teach these topics. To develop an understanding of participants' opinion about the different aspects of instruction, we asked them to describe their perspectives about concepts, design methods, tool use, and practical development skills such as project-based learning. We specifically asked participants to describe aspects of their training that helped them in their current work.

*4.1.3* ***Analysis***. To acquire a more comprehensive picture of interviewees' views on the current needs in VR interaction design education and potential approaches to addressing these needs, the authors closely read through the interview transcriptions. During this process, a set of topics and subtopics were identified which formed the basis for the coding scheme. These codes were then used to iteratively combine and refine elements from the interview data.

### 4.2 Findings

We synthesized the following findings supported by rich evidence from interview analysis and our course meta-review.

***Interaction designers are currently facing a moving target problem.*** One of the key issues raised by 3 of our participants was that VR design education cannot keep up with rapid changes in the VR space. P5 noted that VR is the *"next frontier of computing"* which provides a platform with endless opportunities. However, P1 compared VR to the *"Wild West"*, given that there is still a lack of robust, standardized design guidelines in VR such as those that are common in more established areas like web design. This new frontier presents an exciting opportunity for researchers to explore but it is still difficult to establish a consistent VR design curriculum. Thus, as P1 stated, *"VR design is far from mainstream within university programs today"*. With the understanding that VR is currently an emerging technology, P5 emphasized that this will change. They maintained that students currently see VR as being novel but much like desktop PCs and WIMP GUIs, it will become a standard form of interaction in the next wave of computing.

***Prototyping is currently an underrepresented skill in the VR curriculum.*** In order to help students design and evaluate solutions in a new domain, prototyping is a key activity. It offers designers the ability to communicate ideas quickly with the tools they have available, as described by P6. However, both P3 and P6 indicated that practical prototyping skills receive little coverage in VR curricula. This agrees with the findings from our content analysis. P6 continued by saying students are often obliged to learn prototyping skills outside of their formal training.

The main challenge in supporting VR prototyping is that there is not a smooth transition between low fidelity to high fidelity prototypes. Medium to high fidelity VR prototypes often require programming proficiency and knowledge of 3D frameworks, and there are only a few tools available to support this transition. Indeed, P1 went so far as to say *"I would not be able to do prototyping [in VR] if I didn't learn how to code."* Therefore, it has been challenging to integrate a formal way of transiting between the low-to-high and medium-to-high fidelity prototyping into the VR curriculum.

***Design thinking is largely missing from VR designers' formal education.*** General design thinking methodology is an important approach to interaction design. It allows designers to clearly articulate problems they are trying to address and iteratively refine solutions. Despite the importance of design thinking in understanding user needs, it does not currently receive sufficient coverage in the training available to VR designers based on our course analysis. Moreover, P1, P6, and P7 all maintained that they felt education in design thinking is limited in current university programs and reported they had needed to learn how to apply this methodology after completing their studies. In the case of P6 and P1, both indicated that design thinking was largely missing from their education.

One notable subtopic which emerged in the discussion of design thinking and user-centered design was the balance between education in design methodology and technical development training. Almost all participants indicated that both areas are highly important for VR designers but both P4 and P7 maintained that adequate training in design methods is more important than training in technical development skills.

***Some aspects of professional VR development are not sufficiently covered.*** While some participants indicated that training in design methods is highly important, several others emphasized the value of technical development skills for VR designers. P3 specifically emphasized the need for technical competency to understand development concerns such as maintainability, compatibility with existing systems, and performance for designers in industrial positions. They indicated that they felt this was missing from current university curricula. Moreover, P6 and P1 both indicated that designers without these technical skills would have limited agency to implement their designs.

***Competence in evaluation methods is essential but is not provided by the current curriculum.*** Besides training VR designers in prototyping and design thinking, another key theme that emerged in the interviews was the importance of system evaluation skills. The ability to understand how well a solution meets user needs is crucial for VR interaction design. However, both P3 and P7 expressed concerns that adequate evaluation techniques were missing from the VR designers' education. P7 stated, *"getting actual feedback from people in the real world is something that is not really taught in school."* Moreover, P3 described how, many types of user evaluation methods were not intuitive, and designers were often not adequately trained during their education to perform user studies. This deficiency is significant considering that iterative refinement is unlikely to be successful without properly assessing the efficacy of a solution.

## 5 RECOMMENDATIONS

Based on our course analysis and interview study, it is clear that the education in VR interaction design is very fragmented. While a few courses are covering this emerging area relatively well, there is still not a clear focus on this topic in general. The current curricula do not provide students with sufficient background needed to be successful VR interaction designers. To address the current problems in VR education, we have generated the following recommendations:

***R1: Practical design exercises should be used as a key teaching tool.*** A central recommendation for addressing the deficiencies we identified is to add practical design exercises to the current VR curriculum. These exercises should allow students to apply design thinking, prototyping, and user evaluation as part of their learning objectives. Ultimately, the goals are to enable students to work with closely related real world problems, develop the ability to understand user needs, and iteratively turn ideas into solutions.

To this end, it is important that the exercises teach students what questions to ask to understand user needs. After forming an initial definition of user needs, students should then perform prototyping and user evaluations to iteratively refine their solution and update their understanding of the user requirements by incorporating the results of their evaluations. As expressed by both P3 and P7, observing users (which could be their colleagues) interacting with the prototype gives students a practical insight into design issues in the real world.

***R2: Students should be taught fundamental design principles and guidelines.*** While the focus should be on practical design exercises as the primary pedagogical method, it is important that students should also develop a basic understanding of the techniques they will be applying. Therefore, the design curriculum in VR should cover the basics of design thinking and user-centered design. There should be an emphasis on the principles of prototyping and user evaluation. This is particularly important in solving the moving target problem for interaction design. While tools come and go, and new technologies emerge and evolve quickly, the fundamental principles of design that are applicable to a particular space remain the same. Understanding and applying these principles is key to staying on top. Thus, students should be provided with a strong background in design principles and guidelines.

***R3: Students should be trained in prototyping and ideation tools.*** In order to effectively support prototyping, it is also important that students should receive sufficient training in the required tools that would help them bring ideas to life quickly. This can include simple low-fidelity techniques such as sketching and physical 3D prototyping, as well as high fidelity techniques. In the case of the latter, it is more helpful to train students in high fidelity prototyping technologies such as DraftXR[1] which have a lower barrier of entry in terms of programming compared to 3D frameworks like Unity. This would help students focus on thinking more about the design problem than struggling first with challenges that come with technical implementation.

***R4: Courses should provide students with industry insights.*** Exposure to the real world is a great advantage to any learning endeavor, particularly for students who would like to work in the industry. Only a few of the schools in this study had sessions where industry professionals gave talks about what they were currently building in their organizations. Unlike student internships, experiences gained from such talks are part of in-class pedagogy. Integrating such professional talks into the curriculum would give students a diverse perspective to their training and bridge the gap between their desired future and their current state of learning.

***R5: Education should not be limited to the classroom.*** Learning outside of the confines of the classroom encourages students to solve practical problems. This offers an advantage to the teaching and learning outcomes for both teachers and students. As P6 suggested, *"I would try to tell them to find a problem that they have around them, maybe at their home...and be creative about it"*. Experiential-based learning beyond the classroom is essential as it enforces connections with the real world that are needed *"to recreate either a digital twin of the world or a new spatial experience that is immersive."* (P6)

## 6 LIMITATIONS AND FUTURE WORK

In our approach to selecting institutions, we have endeavored to limit the effect of investigator's bias. We aimed to represent a diverse set of institutions across different communities of learning, including those that are prominent in either research or teaching. However, we acknowledge that our approach still has the potential to exclude certain institutions that may have instruction covering VR interaction design. Also, our course data overrepresents Western institutions, particularly universities from the United States. This may restrict insights into educational needs in other contexts. As a potential future direction, it would be informative to explore a wider range of schools outside of those regarded as top tier and in different regions to investigate the generalizability of our findings and identify additional needs.

Course data was collected from online syllabi and course materials. However, the information provided by some of the courses may have excluded topics that were discussed in class. In future work, interviewing course instructors could highlight instructors' insights and augment our findings with information not provided in the course content and syllabi.

In conclusion, our findings reveal that the current VR curriculum lacks theoretical and practical consideration of interaction design guidelines in VR. Therefore, we see the primary beneficiaries of our study to be course instructors and stakeholders developing educational content in universities and other learning institutions. While it was infeasible to tackle the resource-intensive task of developing a new curriculum within the scope of this study, we hope that our contributions will lead to an improved curriculum for VR interaction design in the future.

---
[1]https://www.draftxr.com

# Appendices

## A  TOPIC TAXONOMY OF VR TOPIC LABELS AND CORRESPONDING SUBTOPICS

| Taxonomy Labels | Subtopics |
| --- | --- |
| Alternative Interaction Modalities | Sound, haptics, haptic devices, multimodal interaction, sound simulation, spatial audio, voice control in VR |
| Graphics | Object modeling, graphics engines, physics, the graphics pipeline, graphics frameworks (OpenGL, Vulkan, DirectX), geometry, rendering, physics engines, physics of motion, photometric registration |
| VR Interface Design | Adaptive interfaces, VR world design, contextual interfaces, developing VR interactions, multimodal interfaces, human-centered interface design in VR, VR interaction design techniques, presence and immersiveness |
| User properties and limitations | Visual perception, motion sickness, the human visual system, human psychological processes, spatial cognition, human sensation and perception, task load, user movement, selection and manipulation |
| Applications of VR | 3D dataviz, architecture design, gaming, therapy, in-car interfaces, 3D geospatial VR, emerging technologies, location Based and context-aware systems, embedded intelligence/smart Objects, immersive video |
| Sociocultural impacts of VR | Ethical concerns, cultural concerns, career pathways, remote collaboration, social interaction |
| Background of VR | History of VR, Introduction to VR systems, Current state of VR, Frontiers of VR |
| Practical development in VR | VR frameworks (Unity, A-Frame, etc.), programming languages, Git, tracking motion and position, inertial measurement, 3D capturing, VR video capture, state machines, game development in VR, processing 3D data |
| VR hardware | 3D HMDs, AR displays, display optics, Direct-View Light Field Displays, sensors, rendering hardware |

## B  PILOT EMAIL

We contacted 22 professionals with industry experience for an interview using a pilot email (14 E-mails, 8 LinkedIn In-mails). We got 5 out of 14 Emails and 2 out of 8 LinkedIn contacts who responded timely and favorably to our request. Our interview was scheduled on Zoom teleconferencing platform. We include a sample of our pilot email here and a list of questions asked during the interview in Appendix C).

Dear X,

Greetings to you! My name is {anonymous identity}, ****

As part of our ongoing *********, we are conducting interviews with experts in the field of Virtual Reality with a focus on Interaction Design. The interviews are essentially gaining your perspectives on what VR curriculum with a focus on Interaction Design topics in Universities should include and how they can prepare potential students for future industrial positions.

Given your background and expertise, we would really be grateful to have the opportunity to conduct an interview with you. The interview should take only 20 minutes and your unique perspective would help us formalize the insights we have developed so far. We would be happy to share an anonymized version of the interview findings with you.

We are looking forward to hearing from you. If you find this interesting, we would be happy to know your preferred days and times and get some advice from you.

Thank you very much for taking the time to read our email and we hope to hear from you.

Best regards
{anonymous identity }

## C  INTERVIEW QUESTIONS

*Hello thanks for joining **** today, ____. I'm {anonymous identity} Now before we start, we would just like to ask if it would be okay to record this interview session? (Response) Thank you.*

*Q1. Which aspects of instruction in VR interaction design do you feel is the most important for students? (options in case they can't remember any: i) Design Process, ii) Design thinking, Ethics and Guidelines iii) Storyboarding and Physical Prototyping iii) Digital Prototyping and iv) Evaluation of VR experiences etc*

*Q2. In carrying out your day-to-day tasks, what tools and methods do you typically use for interaction design in VR?*

*Q3. Do you feel the topics you learned in school gave you enough foundation to perform your day-to-day job when you are designing VR applications?*

*Q4. If you acquired a skill from your current work that you never learned at school, can you describe how you taught yourself this skill?*

*Q5. From the experience you have gained in the work you do currently, what topics would you like to be included in Interaction (VR) design that is taught in schools today?*

*Q6. What practical exercises would be good for students who are potential interaction designers in VR?*

*Q7. Which usability problems in interaction design (VR) do you find the hardest to solve?*

*Q8. Finally, to what extent do you think the current curricula in VR and related courses prepare students to handle the concerns and usability problems specific to interaction design and why?*

*Demographic Q: Approximately, how many years of industrial experience do you have in interaction design (VR)?*